\documentclass[aps,prd,superscriptaddress,nofootinbib,twocolumn,10pt]{revtex4-2}
\usepackage{color,amsmath,amssymb,graphicx,latexsym,lineno}
\usepackage{threeparttable,multirow,txfonts,subfigure,booktabs}

\graphicspath{{./}{fig/}}

\begin{document} 

\title{Evidence of cosmic-ray acceleration up to sub-PeV energies in the supernova remnant IC 443}

\author{Zhen Cao}
\affiliation{Key Laboratory of Particle Astrophysics \& Experimental Physics Division \& Computing Center, Institute of High Energy Physics, Chinese Academy of Sciences, 100049 Beijing, China}
\affiliation{University of Chinese Academy of Sciences, 100049 Beijing, China}
\affiliation{TIANFU Cosmic Ray Research Center, Chengdu, Sichuan,  China}
 
\author{F. Aharonian}
\affiliation{TIANFU Cosmic Ray Research Center, Chengdu, Sichuan,  China}
\affiliation{University of Science and Technology of China, 230026 Hefei, Anhui, China}
\affiliation{Yerevan State University, 1 Alek Manukyan Street, Yerevan 0025, Armenia}
\affiliation{Max-Planck-Institut for Nuclear Physics, P.O. Box 103980, 69029  Heidelberg, Germany}
 
\author{Y.X. Bai}
\affiliation{Key Laboratory of Particle Astrophysics \& Experimental Physics Division \& Computing Center, Institute of High Energy Physics, Chinese Academy of Sciences, 100049 Beijing, China}
\affiliation{TIANFU Cosmic Ray Research Center, Chengdu, Sichuan,  China}
 
\author{Y.W. Bao}
\affiliation{Tsung-Dao Lee Institute \& School of Physics and Astronomy, Shanghai Jiao Tong University, 200240 Shanghai, China}
 
\author{D. Bastieri}
\affiliation{Center for Astrophysics, Guangzhou University, 510006 Guangzhou, Guangdong, China}
 
\author{X.J. Bi}
\affiliation{Key Laboratory of Particle Astrophysics \& Experimental Physics Division \& Computing Center, Institute of High Energy Physics, Chinese Academy of Sciences, 100049 Beijing, China}
\affiliation{University of Chinese Academy of Sciences, 100049 Beijing, China}
\affiliation{TIANFU Cosmic Ray Research Center, Chengdu, Sichuan,  China}
 
\author{Y.J. Bi}
\affiliation{Key Laboratory of Particle Astrophysics \& Experimental Physics Division \& Computing Center, Institute of High Energy Physics, Chinese Academy of Sciences, 100049 Beijing, China}
\affiliation{TIANFU Cosmic Ray Research Center, Chengdu, Sichuan,  China}
 
\author{W. Bian}
\affiliation{Tsung-Dao Lee Institute \& School of Physics and Astronomy, Shanghai Jiao Tong University, 200240 Shanghai, China}
 
\author{A.V. Bukevich}
\affiliation{Institute for Nuclear Research of Russian Academy of Sciences, 117312 Moscow, Russia}
 
\author{C.M. Cai}
\affiliation{School of Physical Science and Technology \&  School of Information Science and Technology, Southwest Jiaotong University, 610031 Chengdu, Sichuan, China}
 
% ...
\collaboration{The LHAASO Collaboration}

\email[E-mail: ]{chenes@ihep.ac.cn; yyguo@pmo.ac.cn; xyhuang@pmo.ac.cn; liusm@swjtu.edu.cn;
yuanq@pmo.ac.cn; zhangyi@pmo.ac.cn; zhd@pmo.ac.cn}

\begin{abstract}
Supernova remnants (SNRs) have been considered as the primary contributors to cosmic rays (CRs) 
in our Galaxy. However, the maximum energy of particles that can be accelerated by shocks of 
SNRs is uncertain, and SNRs' contribution to CRs around PeV energies is unclear. In this study, 
we present observations of high-energy $\gamma$-ray emission from the SNR IC 443 using the Large 
High Altitude Air Shower Observatory (LHAASO). The morphological analysis reveals a pointlike 
source whose location and spectrum are consistent with those of the Fermi-LAT-detected compact 
source with $\pi^0$-decay signature, and a more extended source that is associated with a newly 
discovered Fermi source. The spectrum of the point source can be described by a power-law 
function with an index of $\sim3.0$, extending beyond $\sim 30$ TeV without apparent cutoff. 
Assuming a hadronic origin of the $\gamma$-ray emission, the $95\%$ lower limit of accelerated 
protons reaches about 300 TeV. The extended source might be associated with IC 443, SNR G189.6+3.3 
or the putative pulsar wind nebula CXOU J061705.3+222127, and can be explained by either a hadronic 
or a leptonic model with particles reaching hundreds of TeV. These LHAASO results provide compelling evidence that sub-PeV CRs can be accelerated efficiently by shocks of SNRs.
\end{abstract}

\maketitle

{\it Introduction.} ---
It is widely believed that Galactic sources have the capability to accelerate cosmic rays (CRs) 
up to energies in the knee region, which represents a distinct break in the CR spectrum around
several PeV \cite{2005APh....24....1A,2013FrPhy...8..748G,2008ApJ...678.1165A,2024PhRvL.132m1002C}. 
These sources, known as PeVatrons, remain elusive despite ongoing efforts to identify them.  
Supernova remnants (SNRs), which accelerate energetic particles via the diffusive shock acceleration
mechanism, are considered to be promising candidates for PeVatrons \cite{2024NatAs...8..425D}.
With the nonlinear effect of the diffusive shock acceleration and the possible magnetic field 
amplification, SNRs are also expected to be able to accelerate CRs to PeV energies
\cite{2001RPPh...64..429M,2013MNRAS.431..415B,2024NatAs...8..425D}. Observations of ultra-high-energy
$\gamma$-ray emission from SNRs, particularly those interacting with dense molecular clouds (MC)
\cite{2010ApJ...712.1147J}, are expected to provide direct evidence of whether SNRs can serve as PeVatrons.

Usually three radiation mechanisms exist for understanding $\gamma$-ray emission of SNRs, the hadronic process with $\gamma$-ray photons being produced via decay of neutral pions, the leptonic processes produced by accelerated electrons through the inverse Compton scattering off background photons and bremsstrahlung in the medium. In the ultra-high-energy regime, the $\gamma$-ray production from the inverse Compton scattering is limited by the Klein-Nishina suppression effect. Nonetheless, establishing a robust hadronic interpretation remains challenging due to the limited knowledge about SNRs themselves and their environment parameters. Gamma-ray observations in the sub-GeV band of several SNRs interacting with MCs found evidence of characteristic $\pi^0$-decay spectral bumps \cite{2010ApJ...710L.151T,2011ApJ...742L..30G,2013Sci...339..807A}, making these objects ideal targets for probing hadronic CR acceleration. 

IC 443 is a middle-aged SNR with an estimated age ranging from 3 to 30 thousand years
\cite{2008AJ....135..796L,1988ApJ...335..215P,2008A&A...485..777T}, 
at a distance of approximately 1.5 kpc \cite{1984ApJ...281..658F}. The interaction of IC 443 
with surrounding MCs has been firmly established through the detection of OH maser emission 
\cite{1999ApJ...511..798C,2006ApJ...652.1288H,2008ApJ...683..189H} and various molecular lines
\cite{1986ApJ...302L..63H,1998ApJ...505..286S,2014ApJ...788..122S}. The remnant exhibits a 
double-shell structure in both optical and radio wavelengths \cite{1986A&A...164..193B}. 
In the X-ray band, IC 443 is predominantly characterized by the thermal emission 
\cite{2006ApJ...649..258T}. In terms of $\gamma$-ray emission, a relatively compact source 
with a Gaussian extension (39\% containment) of about $0.17^{\circ}$ and positionally coincides 
with shocked clouds was detected, and the $\pi^0$-decay bump in the sub-GeV spectrum was 
identified \cite{2010ApJ...710L.151T,2013Sci...339..807A}, offering evidence of hadronic CR 
acceleration in the SNR shock. A possible bow-shock pulsar wind nebula (PWN), CXOU J061705.3+222127
was detected in radio and X-ray in the vicinity of the SNR, but its physical connection with IC
443 SNR has not been well established and the pulsation of the hypothetical pulsar is also not 
found yet \cite{2001ApJ...554L.205O,2004AJ....127.2277L}. Recently, an extended source with bigger
extension ($R_{39}=0.64^{\circ}$) overlapped with IC 443 has been reported \citep{2024arXiv241107162A}, 
but no firm association has been identified. There exists a possible counterpart, SNR G189.6+3.3 
with an extension of about 0.75 degrees, as detected by X-ray observations
\cite{1994A&A...284..573A,2023A&A...680A..83C}. The age of SNR G189.6+3.3 was estimated
to be about $10^5$ yr, and the distance is similar to that of IC 443 ($\sim1.5$ kpc)
\cite{1994A&A...284..573A}. The centroids of the Fermi-LAT extended source and SNR G189.6+3.3 
differ by about $0.3^{\circ}$, which is within the source extensions. TeV emission from IC 443 
was detected by MAGIC \cite{2007ApJ...664L..87A}, VERITAS \cite{2009ApJ...698L.133A}, and
HAWC \cite{2025ApJ...992...22A}. MAGIC identified a point-like source coincident with the densest 
part of the MCs and the position of the 1720 MHz OH maser, indicating a potential hadronic origin 
for the emission \cite{2007ApJ...664L..87A}. VERITAS observed an extended source, with the centroid 
and extension being consistent with the Fermi-LAT small source \cite{2009ApJ...698L.133A}. HAWC
identified two sources in the IC 443 region, a point source positionally consistent with that 
detected by MAGIC and VERITAS, and the other extended one with the centroid near the pulsar
B0611+22 and was postulated to be a pulsar halo \cite{2025ApJ...992...22A}. HAWC did not find 
spectral cutoff of the point source component and inferred that protons up to 65 TeV can be 
accelerated by the SNR.

\begin{figure*}[!htb]
\begin{center}
\includegraphics[width=0.48\textwidth]{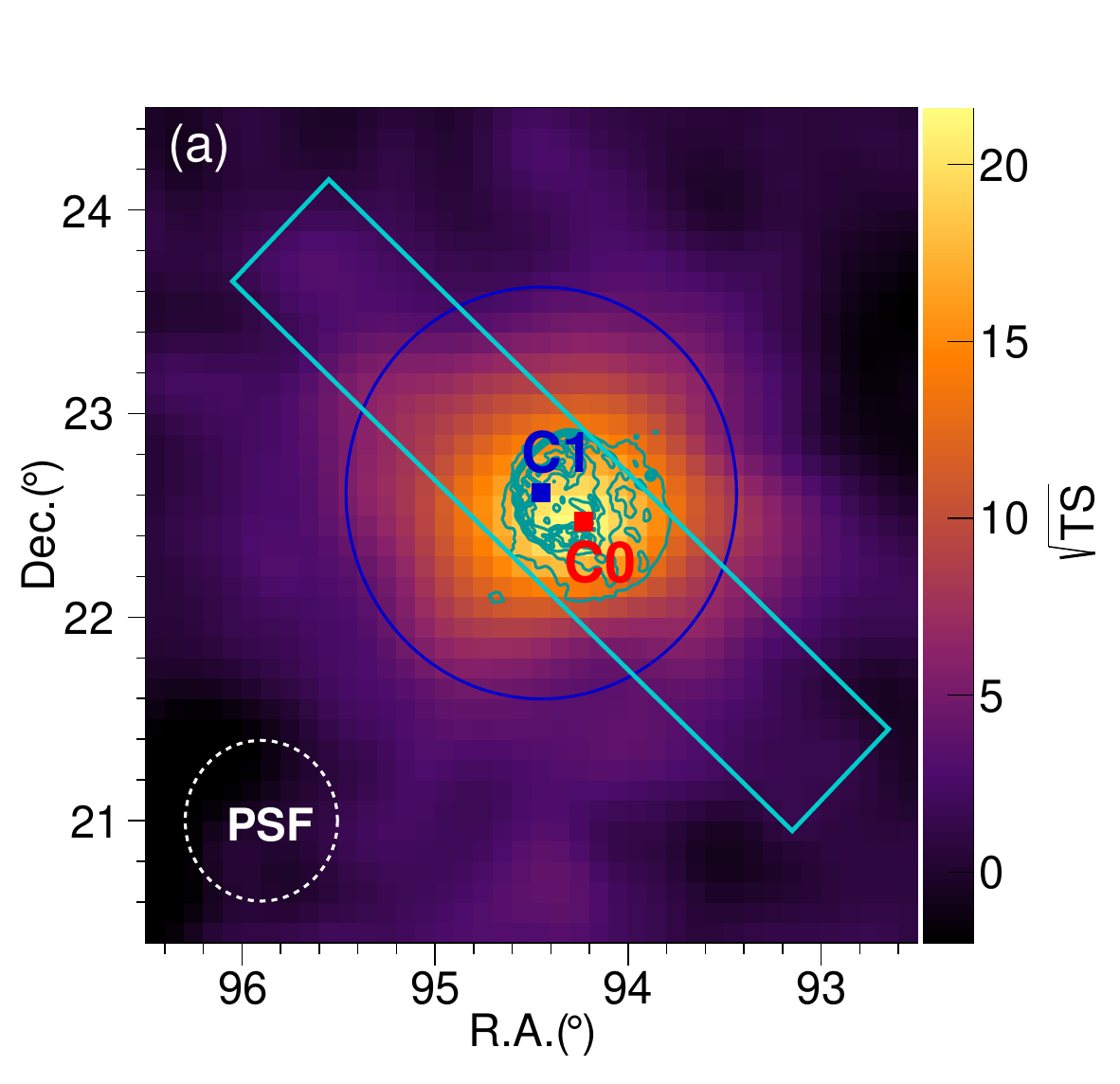}
\includegraphics[width=0.48\textwidth]{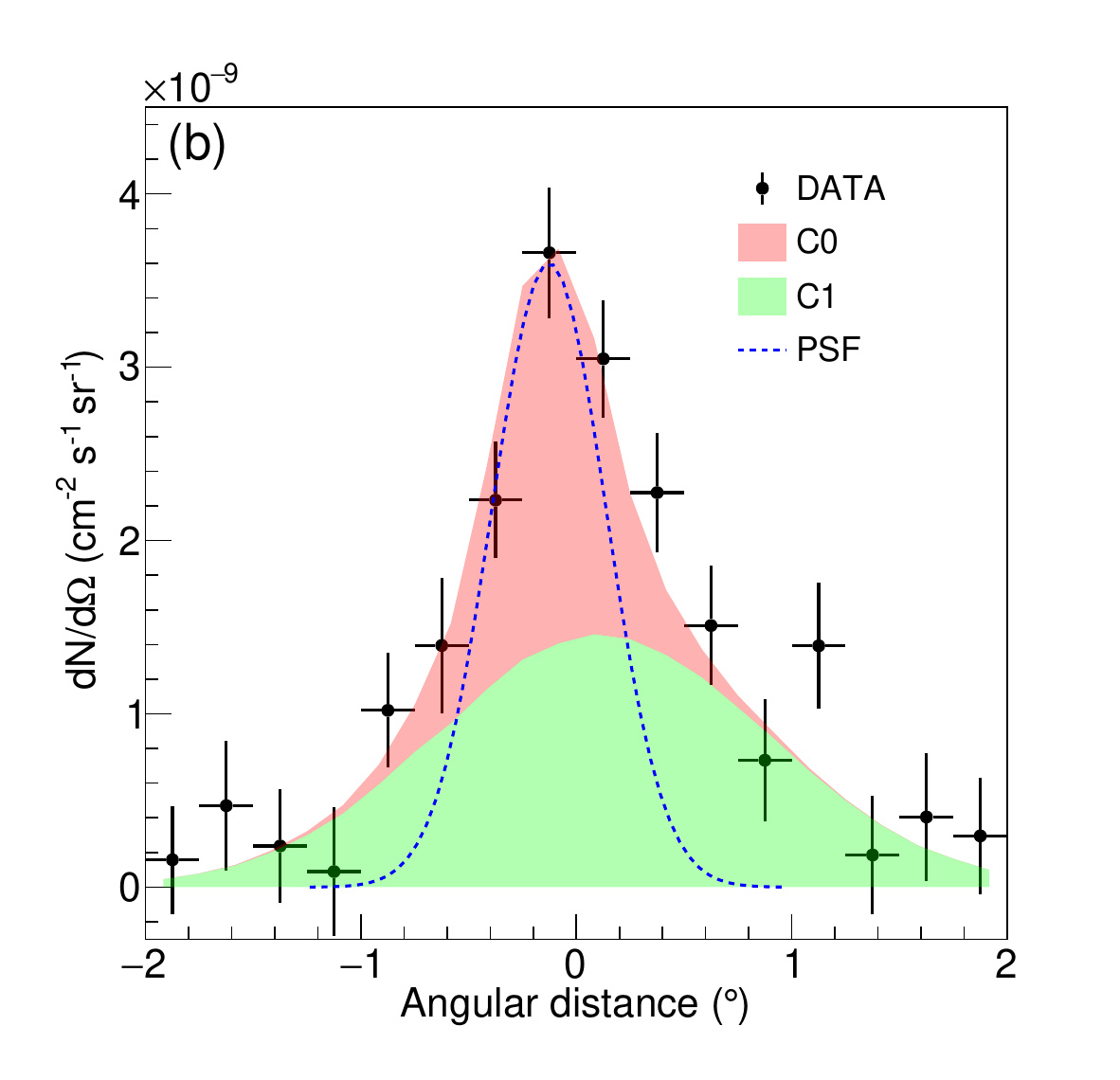}
\includegraphics[width=0.48\textwidth]{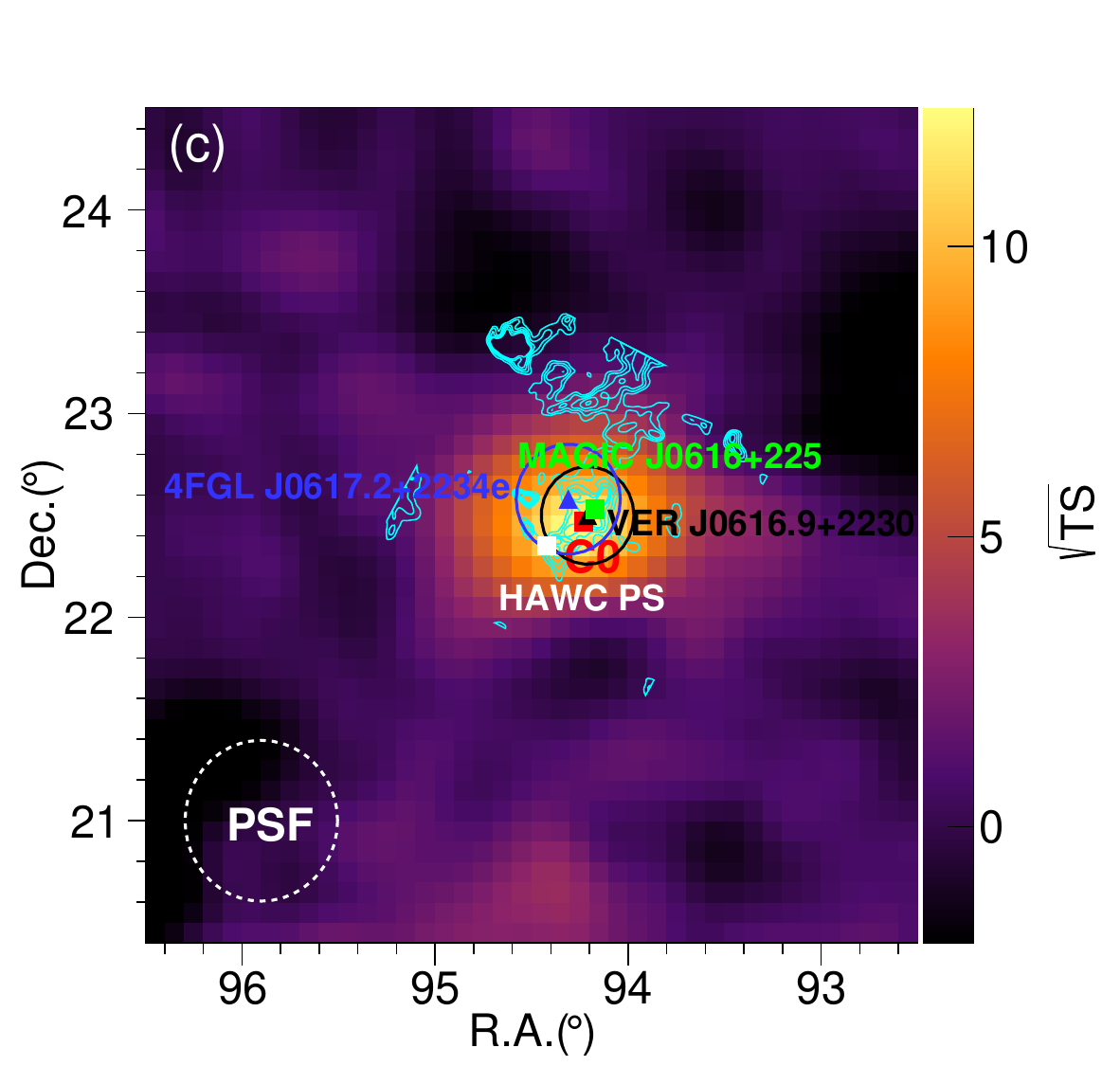}
\includegraphics[width=0.48\textwidth]{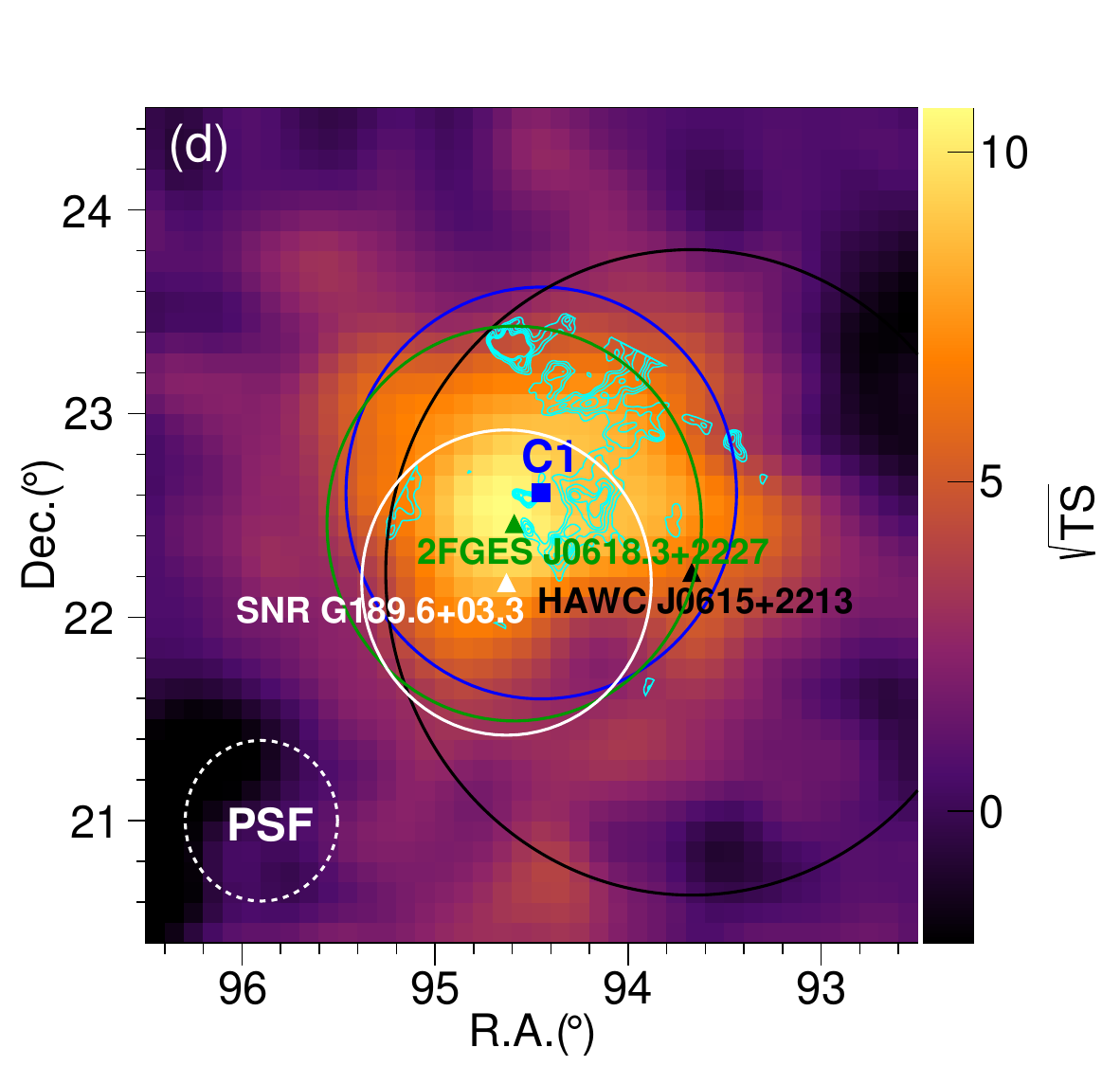}
\caption{Panel (a) shows the LHAASO observed significance map of a $3\times3$ deg$^2$ region 
surrounding IC 443 for $E>0.5$ TeV, overlaid with the 1.4 GHz radio continuum contours in dark-green 
\cite{2008AJ....135..796L}. The dashed line in the bottom right corner shows the diameter of the 
PSF, weighted by the TS value of each energy bin. Red square labels the centroid of C0, and 
blue square and circle show the centroid and $68\%$ containment size (intrinsic) of C1. 
Panel (b) shows the one-dimensional distribution of the integrated $\gamma$-ray fluxes from 
the rectangle box shown in panel (a), together with the PSF convolved profiles of C0 and C1. 
The zero point is chosen as the midpoint between C0 and C1. The dotted line shows the PSF profile 
centered at C0. Panels (c) and (d) show the significance maps of C0 and C1 
components, respectively. The centroids and $68\%$ extensions (if any) observed by other experiments 
are also shown for comparison. The cyan contours depict the shocked molecular gas distribution 
measured by the MWISP project with velocities ranging from $-10$ km/s to $10$ km/s. In panel (d), 
the centroid and extension of SNR G189.6+3.3 are shown in white.
}
\label{fig:sig_map}
\end{center}
\end{figure*} 

\begin{table*}[!ht]
\centering
\footnotesize
\begin{tabular}{l|c|ccccccc} \hline\hline
Model & TS & Name &R.A. ($^{\circ}$) & Dec. ($^{\circ}$) & $R_{39}$ ($^{\circ}$) & $\phi_0$ ($10^{-14}~\rm{TeV^{-1}cm^{-2}s^{-1}}$) & $\alpha$ & $E_{\rm{cut}}~(\rm{TeV})$\\ \hline
Two &\multirow{2}{*}{5630.5} & C0 & 94.27$\pm$0.03 & 22.44$\pm$0.02 & Point & 3.51$\pm$0.43 & 2.95$\pm$0.07 & - \\
\cline{3-9}
Components & & C1 & 94.45$\pm$0.07 & 22.61$\pm$0.06 & 0.67$\pm$0.07 & 17.20$\pm$2.74 & 2.53$\pm$0.14 & 19.65 $\pm$ 8.67 \\  \hline
One Component & 5581.5 & ... & 94.33$\pm$0.03 & 22.52$\pm$0.02 & 0.38$\pm$0.03 & 15.13$\pm$1.56 & 2.68$\pm$0.08 & 34.22$\pm$14.62 \\  \hline \hline
\end{tabular}
\caption{Fitting results of centroids, extensions ($39\%$ containment), fluxes at 3 TeV, and
spectral indices of different components for IC 443.}
\label{tab:rst}
\end{table*}

{\it LHAASO observation and data analysis.} ---
LHAASO is a ground-based extensive air shower experiment located at Haizi Mountain in China, 
with an average altitude of 4410 meters \cite{2022ChPhC..46c0001M}. This hybrid array comprises 
the Kilometer Square Array (KM2A), the Water Cherenkov Detector Array (WCDA), and the Wide 
Field-of-view Cherenkov Telescope Array (WFCTA). The KM2A covers an area of 1.3 km$^2$, and
serves as the most sensitive $\gamma$-ray detector above 20 TeV. The WCDA covers a physical area of 0.08 km$^2$, and can detect $\gamma$ rays down to sub-TeV range. Both KM2A and WCDA 
arrays has a wide field-of-view of approximately $2$ sr, making them well-suited for observing
extended sources. The combination of these two arrays allow us to conduct detailed studies of
$\gamma$-ray sources in a broad energy range. 

This work uses the events collected by the WCDA, from March 5, 2021 to July 31, 2024, 
with a livetime of $\sim$1136 days, and the KM2A, from July 20, 2021, to December 31, 2024,
with a livetime of $\sim1228$ days. We adopt the same selections as described in
Ref.~\cite{2024ApJS..271...25C} to select candidate events. The events are binned with 
0.1$^\circ \times$0.1$^\circ$ grids to make the skymap. The ``direct integration method'' 
\cite{2004ApJ...603..355F} is adopted to calculate the background. 

IC 443 is about 6 degrees away from the Geminga pulsar, and the large extended Geminga halo 
\cite{2017Sci...358..911A} may affect the analysis of IC 443. To properly account for this, 
the region of interest (ROI) is defined as a fan-shape region centered on Geminga pulsar, 
with a radius of 10 degrees and an opening angle of 90 degrees containing IC 443 in its center
(see Fig.~S1 in the {\tt Supplemental Material} which includes
Refs.~\cite{2013arXiv1303.3514A,2020ApJS..247...33A,2023arXiv230712546B}). Within the ROI, 
a diffusion template as 
$f(\theta)\propto \frac{1}{\theta_d(\theta+0.113\theta_d)}e^{[-(\theta/\theta_d)^{1.52}]}$
(adapted from \cite{2021PhRvL.126x1103A}) is adopted to describe Geminga halo emission, where 
$\theta$ is the angular distance from Geminga pulsar and $\theta_d$ is the characteristic 
diffusion width. Note that the Geminga halo exhibits an asymmetric morphology (to be 
published elsewhere), which has been taken into account in the current analysis. Nevertheless, 
within our chosen ROI, considering or neglecting this asymmetry leads to only minor differences 
in the results. The diffuse $\gamma$-ray emission
\cite{2023PhRvL.131o1001C,2025PhRvL.134h1002C} is modelled using the gas template as traced by 
the PLANCK dust opacity \cite{2016A&A...596A.109P} and a broken power-law spectrum. The gas 
template from gas surveys \cite{2016A&A...594A.116H,2022ApJS..262....5D,1991Natur.354..121C} 
is employed as a systematic uncertainty check.

The 3D-likelihood method is employed to simultaneously fit the morphology and spectrum of the
relevant sources in the ROI, which include the target source IC 443, the Geminga halo, and the
diffuse emission in our case. The test statistic (TS) is defined as TS~$=2 \ln ({\mathcal{L}}/{\mathcal{L}_0})$, where $\mathcal{L}_0$ is the maximum likelihood value for the 
null hypothesis and $\mathcal{L}$ is the maximum likelihood for the source hypothesis.

{\it Results.} ---
The significance ($\sqrt{\rm TS}$) map of a $3\times3$ deg$^2$ region centered at IC 443 for 
$E>0.5$ TeV derived with the LHAASO data, calculated for each 0.1$^\circ \times$0.1$^\circ$ 
pixel assuming a point source in the pixel after subtracting the Geminga halo and the diffuse 
background, is shown in panel (a) of Fig.~\ref{fig:sig_map}. Bright extended excess emission 
around IC 443 can be detected. Assuming a Gaussian morphology and an exponentially cutoff 
power-law (ECPL) spectrum, $\phi(E)=\phi_0(E/{3~\text{TeV}})^{-\alpha}e^{-E/E_{\rm{cut}}}$, 
of the emission, it has been found that an extended source with a total significance 
(for $E\gtrsim 0.5$ TeV) of $\sim 26 \sigma$ is detected. The intrinsic extension 
($39\%$ containment) of the source is found to be $R_{39}=0.38^{\circ}\pm0.03^{\circ}$, which 
is between the Fermi-LAT detected compact source and the more extended one. 

We thus test whether the emission can be separated into two components. Via adding one 
more Gaussian template with a power-law (PL) spectrum, $\phi=\phi_0(E/3~{\rm TeV})^{-\alpha}$,
we find that the overall TS value of the fitting increases by about 49 compared with the 
one-source hypothesis. Given 5 more free parameters, it means that the other source component 
is favored with a significance of $6.0\sigma$. The fitting results of different components are 
given in Table \ref{tab:rst}. In the two-component hypothesis, the compact one (C0) is found to 
be a pointlike source with a significance of $10.5\sigma$ and the $95\%$ upper limit of the 
extension being $0.27^{\circ}$, and the extended one (C1) has a significance of $13.1\sigma$ 
and an extension of $R_{39}=0.67^{\circ}\pm0.07^{\circ}$. 
The one-dimensional distribution of the integrated $\gamma$-ray fluxes from the rectangle 
box region labelled in panel (a), together with the profiles of C0 and C1 convolved with the point 
spread function (PSF), is given in panel (b). The zero point is chosen as the midpoint between 
C0 and C1. The dotted line shows the PSF profile centered at C0. This plot indicates that the 
total emission can indeed be decomposed into a pointlike component and an extended component.
In panels (c) and (d) of Fig.~\ref{fig:sig_map}, we show the significance 
maps of C0 and C1 components, respectively, compared with results measured by other experiments. 
Source C1 has been subtracted when producing the significance map of C0, and vise versa.
The LHAASO observed centroids and extensions of both components are in good agreement with those 
detected by Fermi-LAT, as well as MAGIC and VERITAS. While source C0 is roughly consistent with 
the point source observed by HAWC, the centroid and extension of C1 are different from the HAWC 
extended source. The LHAASO source C1 overlaps with SNR G189.6+3.3, with centroids differing by 
about $0.47^{\circ}$.

Green contours in panels (c) and (d) of Fig.~\ref{fig:sig_map} show the molecular gas distribution
around IC 443 as traced by the CO emission with velocities 
ranging from $-10$ km/s to $10$ km/s, observed by the Milky Way Imaging Scroll Painting 
(MWISP; \cite{2019ApJS..240....9S}) project. 
It has been shown that the molecular gas content within this velocity range constitutes the majority of the line-of-sight gas, with only a few high velocity dense clumps due possibly to the SNR shock \cite{2019ApJS..240....9S}.
There is strong evidence that interactions between 
the SNR shock and molecular clouds existing in the IC 443 region, such as the OH maser, line broadening 
and so on \cite{1986ApJ...302L..63H,2008ApJ...683..189H}, indicating that the gas is located at similar distance with the SNR. Apart from shocked clouds in the vicinity 
of compact source C0 \cite{1986ApJ...302L..63H}, there is extended molecular mass distribution 
in a wider region around source C1 (see panel (d) of Fig.~\ref{fig:sig_map}). The results 
indicate that both sources may be produced by hadronic interactions between accelerated protons 
and the molecular gas.

\begin{figure}[!htb]
\begin{center}
\includegraphics[width=0.48\textwidth]{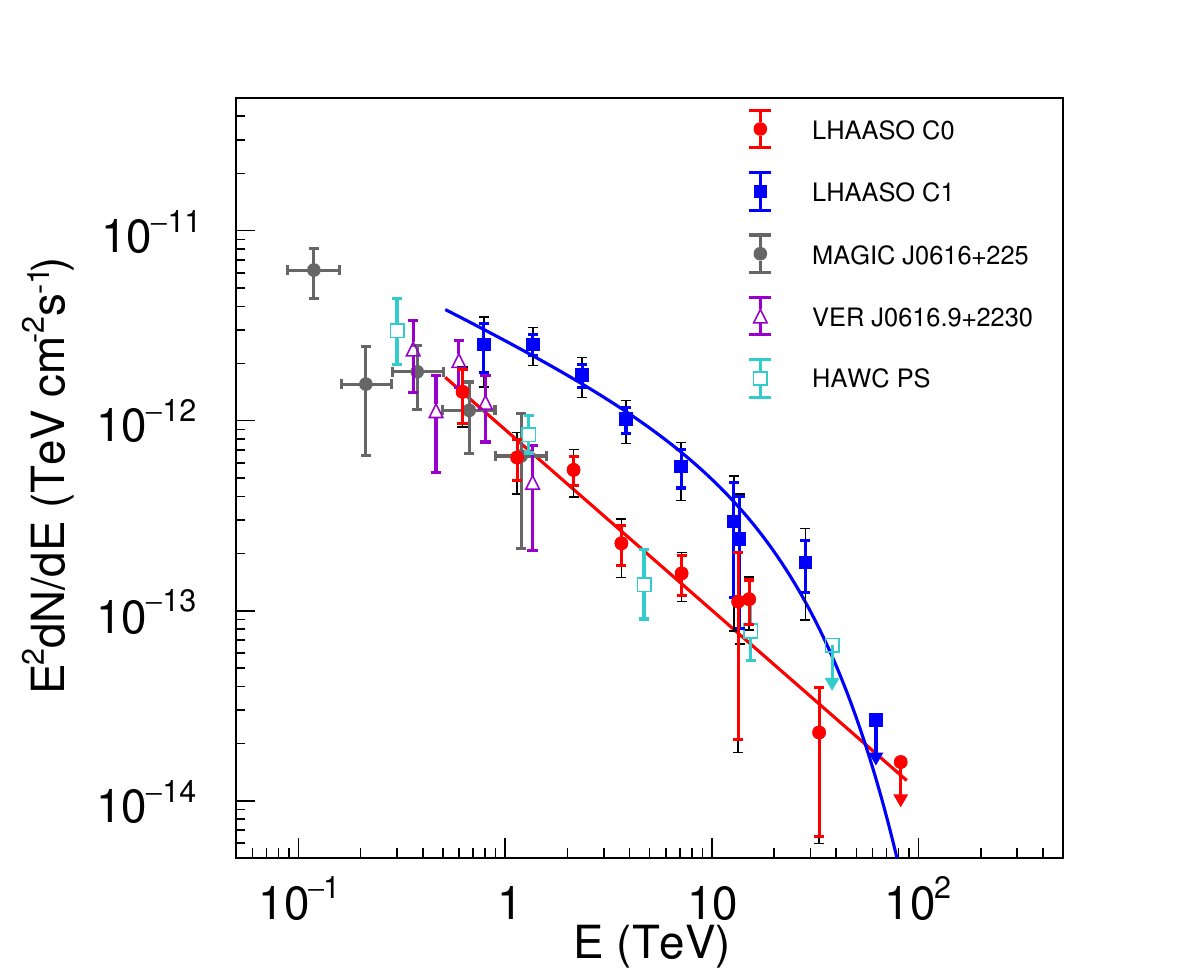}
\caption{The SEDs of the two sources C0 and C1, with statistical errors (red) and total errors
including statistical and systematic ones (black). Arrows show the 95\% confidence level upper
limits, and solid lines show the best-fitting spectra of the two sources. Results measured by 
MAGIC \cite{2007ApJ...664L..87A}, VERITAS \cite{2009ApJ...698L.133A}, and HAWC 
\cite{2025ApJ...992...22A} are also shown. 
}
\label{fig:SED_LHAASO}
\end{center}
\end{figure}

A PL form is found to well describe the spectrum of C0. Fitting with an ECPL function results an 
increase of the TS value of $3.7$, corresponding to a significance of $1.9\sigma$. For C1 component, 
the spectrum fitting favors an ECPL function, with a cutoff significance of $\sim 5.1\sigma$. 
The derived cutoff energy for C1 component is $19.65 \pm 8.67$ TeV, which shows degeneracy with 
the spectral index. Fitting results of the spectral parameters are given in Table
\ref{tab:rst}. The spectral energy distributions (SED) of these two components are shown in
Fig.~\ref{fig:SED_LHAASO}. The SED data can be found in Table~S2 in the 
{\tt Supplemental Material}. Measurements of VHE emission by other experiments are also shown 
for comparison. The LHAASO SED of C0 is consistent with previous measurements, but extend to 
higher energies.

{\it Systematic uncertainties.} ---
The systematic uncertainty in the source location is primarily attributed to the pointing error, 
which is approximately $0.04^{\circ}$ for WCDA and $0.03^{\circ}$ for KM2A \cite{2024ApJS..271...25C}. 
The systematic uncertainty on the source size is mainly due to the uncertainties of the PSF, and is
estimated to be about $0.05^{\circ}$ (for $R_{39}$) for WCDA and $0.08^{\circ}$ (for $R_{39}$) for 
KM2A \cite{2024ApJS..271...25C}. Regarding the flux measurements, the systematic uncertainties on 
the absolute flux are estimated to be about $9.6\%$ for WCDA \cite{2023Sci...380.1390L} and $7\%$ 
for KM2A \cite{2021ChPhC..45b5002A}, due mainly to various kinds of model assumptions of the Monte 
Carlo simulation. The uncertainties of the diffuse $\gamma$-ray background and Geminga halo would 
affect the measurements of IC 433. Comparison of the results between the fittings assuming a fixed 
diffuse background \cite{2025PhRvL.134h1002C} and a free diffuse background gives only slight 
impacts on the flux measurements of C1. Using the gas map from gas surveys as diffuse template
results in very minor changes of the results of both C0 and C1.
While the detailed analysis of the Geminga halo will be published elsewhere, we study the 
impact on IC 443 due to the uncertainty of the morphology assumption of the Geminga halo, 
and find again the main impacts are on fluxes of source C1 for $E>20$ TeV. 
See Fig.~S2 of the {\tt Supplemental Material} for more details. 
The total systematic uncertainties on the SEDs are added in quadrature to the statistical ones
and are shown by black errorbars in Fig.~\ref{fig:SED_LHAASO}.

\begin{figure}[!htb]
\begin{center}
\includegraphics[width=0.48\textwidth]{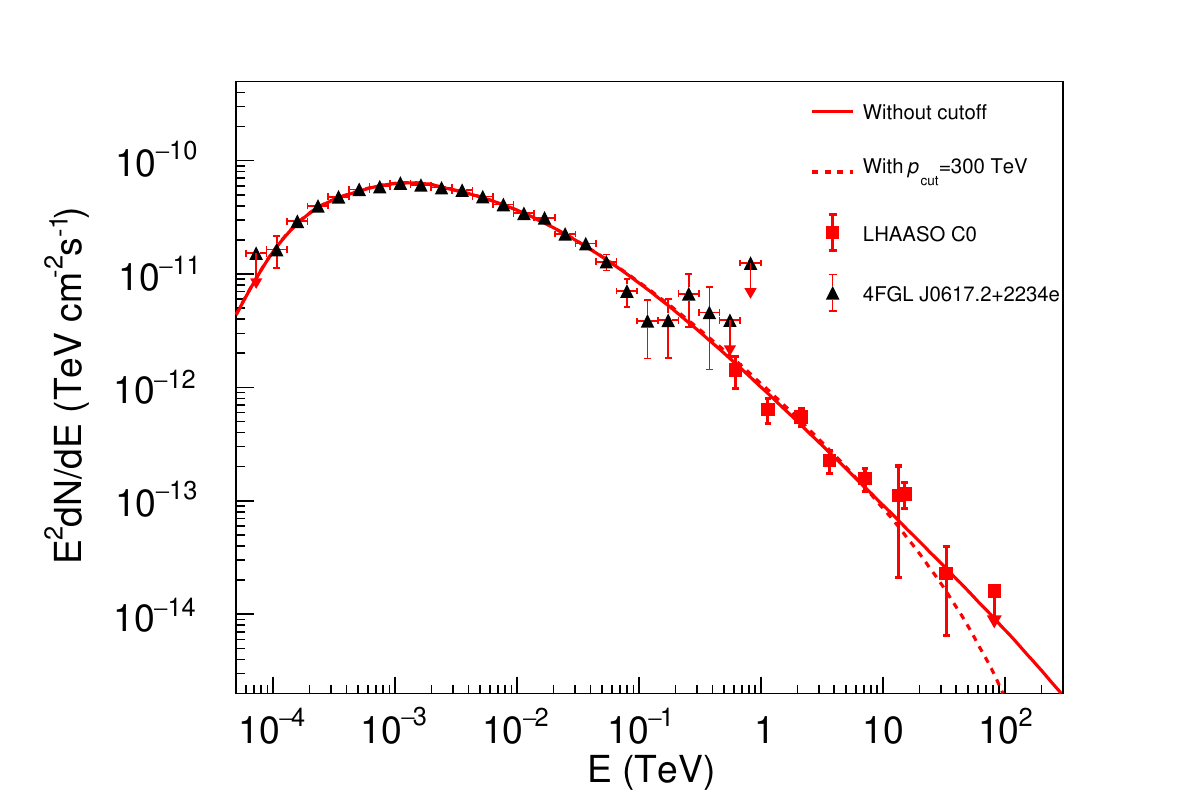}
\includegraphics[width=0.48\textwidth]{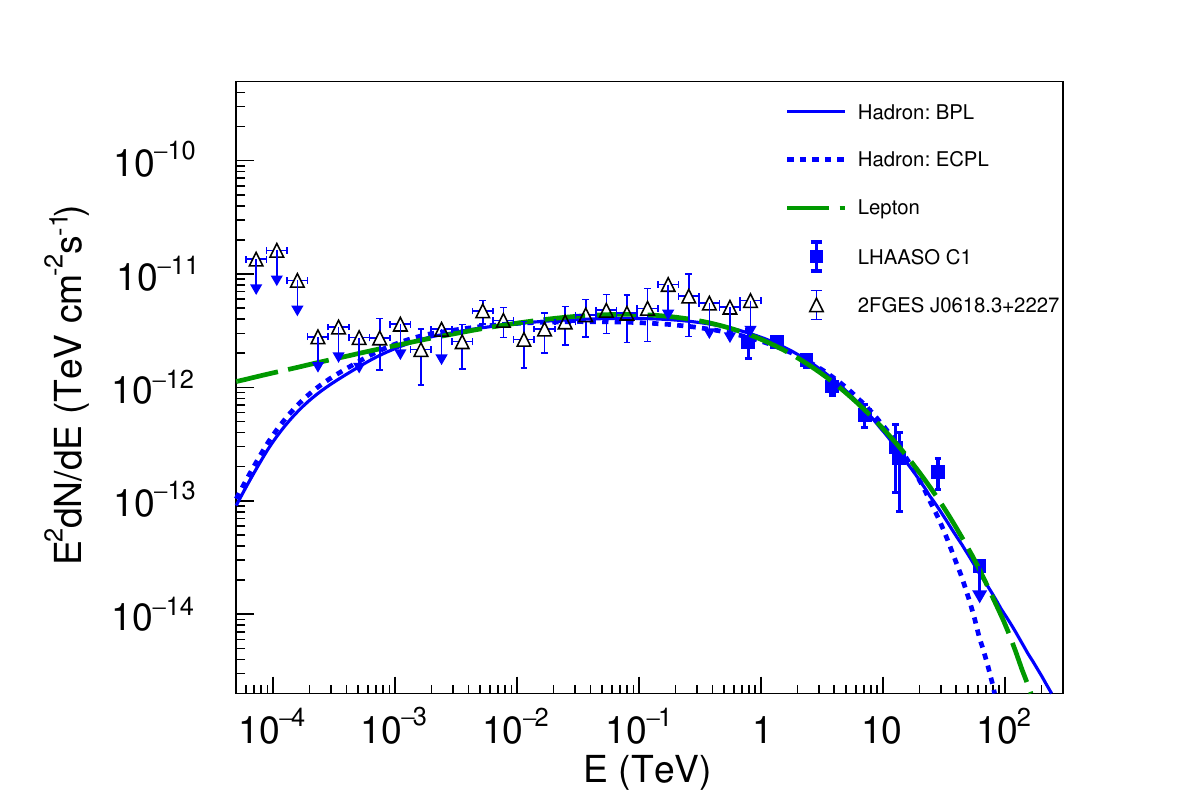}
\caption{Gamma-ray spectra of C0 (top panel) and C1 (bottom panel) as measured by Fermi-LAT 
and LHAASO. Solid lines in the plots show the hadronic model predictions of the spectra with
Eq. (1), assuming no spectral cutoff of protons ($p_{\rm cut}=\infty$). In the top panel, 
the dashed line shows the hadronic model flux for $p_{\rm cut}=300$ TeV (for C0), 
and in the bottom panel, the dashed and long-dashed lines show predictions of 
the hadronic ECPL and the leptonic model (for C1), respectively.}
\label{fig:SED_explain}
\end{center}
\end{figure}

{\it Discussion.} ---
The LHAASO source C0 is morphologically consistent with the Fermi-LAT compact source with 
$\pi^0$-decay signature. The spectrum also nicely connect with that of Fermi-LAT (see Table~S3 
of the {\tt Supplemental Material} for the re-analysis of Fermi-LAT data), suggesting that it 
is very likely to be the high energy counterpart of the Fermi-LAT source. The wide-band 
$\gamma$-ray emission can be modelled with a hadronic model. The proton spectrum around the 
SNR is parameterized as a broken power-law distribution with an exponential cutoff  
\begin{equation}
Q(p)=Q_0 p^{-s_1}\left[1+(p/p_{\rm br})^{s_2-s_1}\right]^{-1} {\rm e}^{-p/p_{\rm cut}},
\label{eq:parent_sp}
\end{equation}
where $p$, $p_{\rm br}$, and $p_{\rm cut}$ are the momentum, break momentum, and cutoff momentum
of protons, $s_1$ and $s_2$ are spectral indices below and after $p_{\rm br}$. 
The spectral break is required to fit the Fermi-LAT data \cite{2013Sci...339..807A}.
Note that the parameter $p_{\rm cut}$ is a characteristic number to describe the spectral
behavior of protons at the highest end, and may not directly correspond to the cutoff energy of 
$\gamma$-ray photons. Using the $\gamma$-ray yield parameterization of Ref.~\cite{2023CoPhC.28708698K}, 
we obtain the expected $\gamma$-ray spectra as shown in the top panel of Fig.~\ref{fig:SED_explain}. 
Here we assume an average gas density of 20 cm$^{-3}$ \cite{2013Sci...339..807A}. 
Since no significant spectral cutoff of the LHAASO spectrum of C0 is found, we assume 
$p_{\rm cut}=\infty$, and get the proton spectrum parameters as: $s_1=2.28_{-0.02}^{+0.03}$,
$s_2=3.13_{-0.06}^{+0.05}$, $p_{\rm br}=0.38_{-0.13}^{+0.16}$ TeV, and 
$W_p=5.67_{-0.57}^{+0.57} \times 10^{49}(n/20~{\rm cm}^{-3})^{-1}$ erg which is the total energy 
of protons with kinetic energy above 1 GeV. Comparison of the model fitting spectrum with the 
measurements by Fermi-LAT and LHAASO is shown in the top panel of Fig.~\ref{fig:SED_explain}. 

\begin{figure}[!htb]
\centering
\includegraphics[width=0.48\textwidth]{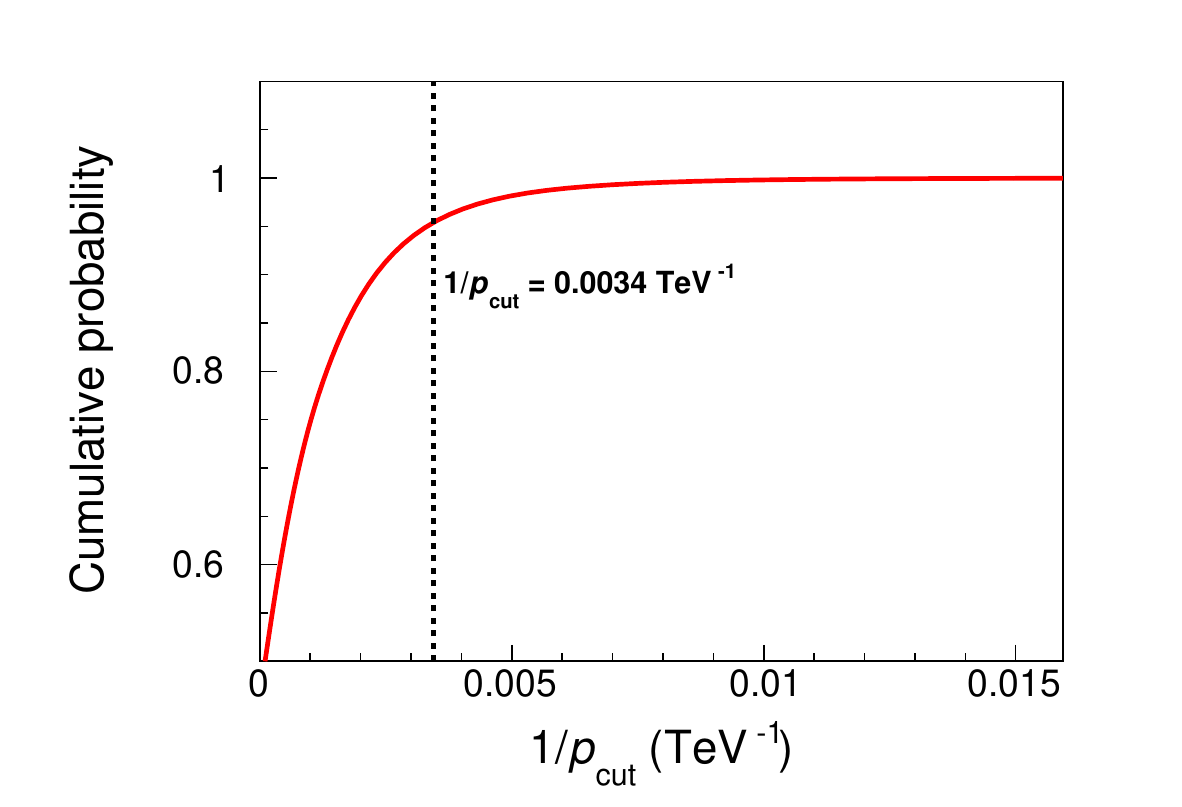}
\caption{Cumulative probability distribution of parameter $1/p_{\rm cut}$ for source C0. The $95\%$ 
lower limit on $1/p_{\rm cut}$ is 0.0034 TeV$^{-1}$, as indicated by the vertical dashed line, 
corresponding to $p_{\rm cut} \approx 300$ TeV. }
\label{fig:pcut1D}
\end{figure}

We can derive a constraint on the cutoff energy of accelerated protons for source C0. 
Fig.~\ref{fig:pcut1D} shows the probability distribution of the inverse of cutoff parameter,
$1/p_{\rm cut}$. Both the Fermi-LAT data and the LHAASO data are included in the likelihood 
computation, and the other spectral parameters are left free to be optimized in the calculation.
The $95\%$ upper limit\footnote{Note that, the probability distribution of $1/p_{\rm cut}$ is
single-sided since we restrict $1/p_{\rm cut}$ to be positive. There is one half probability 
that the statistical fluctuation gives negative result of $1/p_{\rm cut}$ is eliminated
\cite{Chernoff:1954eli}. We thus add this probability to $1/p_{\rm cut}=0$ and integrate to 
a cumulative probability of 0.95.} of $1/p_{\rm cut}$ is found to be about 0.0034 TeV$^{-1}$, 
as labelled by the vertical line. This corresponds to a lower limit of $p_{\rm cut}\approx 300$ 
TeV. As a comparison, we also show the model curve with $p_{\rm cut}=300$ TeV by the dashed 
line in Fig.~\ref{fig:SED_explain}.

The standard diffusive shock acceleration mechanism of SNRs predicts that the highest energy
particles are accelerated near the transition from free-expansion to Sedov phase of the shock evolution. 
The lower limit of proton acceleration of $\sim300$ TeV challenges the traditional acceleration model
prediction of tens of TeV for the test particle approximation \cite{1983A&A...125..249L}, and may imply
 magnetic field amplification and turbulence generation in the shock region \cite{2004MNRAS.353..550B},
nonlinear effects during the shock evolution (e.g., \cite{1996APh.....5..367B}), or specific 
configuration of magnetic fields around the shock \cite{1987ApJ...313..842J}. This result 
distinguishes IC 443 from some young SNRs like Cas A, where a spectral bump followed by a quick flux
decrease in the TeV band is observed \cite{2025ApJ...982L..33C}. The TeV bump of Cas A may be due to 
the leptonic process and is a reflection of the radiative energy loss effects on the electron 
acceleration limit, or due to the hadronic particle acceleration limit of a young SNR at the current 
stage \cite{2025ApJ...982L..33C}. The prominent very high energy emission from IC 443 likely benefits 
from its higher ambient density, which enhances the $pp$ collision efficiency and reveals the 
cumulative history of particle acceleration. The 0.38 TeV break is possibly due to the effect that
high-energy particles become less effectively confined and diffuse out from the shock region 
\cite{2011NatCo...2..194M}.

The LHAASO source C1 could be the counterpart of Fermi-LAT extended source, and may be related 
to IC 443, SNR G189.6+3.3, or CXOU J061705.3+222127. The broadband SED from GeV to 100 TeV for 
source C1 is distinct from that of C0. Assuming also a hadronic emission mechanism, we obtain the 
proton spectrum parameters as: $s_1=1.94_{-0.10}^{+0.06}$, $s_2=3.78_{-0.28}^{+0.46}$, 
$p_{\rm br}=22.6_{-8.3}^{+13.6}$ TeV, and 
$W_p=8.18_{-0.83}^{+0.85} \times 10^{49}(n/1~{\rm cm}^{-3})^{-1}$ erg. 
Note that, although using the ECPL model to fit the LHAASO data favors a spectral cutoff 
of C1, its wide-band spectral behavior affects the fitting result. When using Eq. (1) to describe 
the proton spectrum, we find that the spectral cutoff of C1 is insignificant (the TS value 
increases by about 1.24 compared with infinite cutoff). 
The solid line in the bottom panel of Fig.~\ref{fig:SED_explain} shows the hadronic model 
fitting result for $p_{\rm cut}=\infty$, corresponding to a broken power-law (BPL) model. 

For source C1, we also test the ECPL model for proton spectrum across a wide $\gamma$-ray band for
comparison, and find that the current data cannot distinguish these two models (The TS value for 
the ECPL model decreases by 0.8). The fitting parameters of the ECPL model are: $s=1.98 \pm 0.06$, 
$p_{\rm cut}=41.3 \pm 12.7$ TeV, and $W_p=(8.28 \pm 0.85) \times 10^{49}(n/1~{\rm cm}^{-3})^{-1}$ erg.
The apparent proton spectrum for C1 is different from that of C0. In particular, the break energy of 
$\sim 22$ TeV (if described by the BPL form) is notably higher than that of C0 ($\sim 0.38$ TeV). 
If source C1 is associated with SNR G189.6+3.3, this may imply that SNR G189.6+3.3 has a higher 
shock speed and hence a higher maximum acceleration limit than IC 443 at present. The X-ray 
observation of G189.6+3.3 gives an electron temperature on average higher than that of IC 443
\cite{2023A&A...680A..83C,2006ApJ...649..258T}, likely supports this hypothesis although 
G189.6+3.3 might be older. If sources C0 and C1 both originate from IC 443, the spectral 
difference may be attributed to propagation effects of particles, which results in a suppression 
of low-energy particles due to inefficient propagation for the extended source 
\cite{2007ApJ...665L.131G, 2020pesr.book.....V}. Note, however, although there are molecular 
gases in the extension region of source C1, the morphology of C1 does not show clear correlation 
with the gas distribution. 

Alternatively, a leptonic scenario with the inverse Compton scattering emission of accelerated 
electrons may also explain these measurements, as shown by the dashed line in the bottom panel of
Fig.~\ref{fig:SED_explain}. Here the background radiation fields are approximated with two gray body
components including the cosmic microwave background with a temperature of 2.725 K and an energy 
density of 0.26 eV~cm$^{-3}$, an infrared background with a temperature of 30 K and an energy density 
of 1.0 eV~cm$^{-3}$ \cite{2013Sci...339..807A}. Parameters for electrons are: $s_1=2.49_{-1.12}^{+0.40}$,
$s_2=3.84_{-0.47}^{+1.12}$, $p_{\rm br}=3.66_{-3.00}^{+16.08}$ TeV, $W_e=2.12_{-1.96}^{+12.04} 
\times 10^{48}$ erg and a fixed cutoff energy $p_{\rm cut}=125$ TeV, which corresponding to the 
cooling energy of electrons in the above background photon fields and a $3~\mu$G magnetic field 
for an age of $\sim10$ kyr (IC 443). The effective propagation distance \cite{1995PhRvD..52.3265A} 
is about $\sqrt{2Dt}\approx 17$ pc for a slow diffusion coefficient \cite{2017Sci...358..911A}, 
which is also consistent with the extension of C1 (the $39\%$ containment radius of $\sim17.5$ pc). 
However, if source C1 is associated with SNR G189.6+3.3, the parameters will be different from 
the above estimate. Another possibility of C1 is the halo emission associated with the PWN CXOU
J061705.3+222127, although the position of the PWN deviates from the centroid of C1 by about 0.3 
degrees and the age of the PWN seems to be somehow young. At present it is difficult to judge 
which one explains the data better than the other, and we need additional multi-wavelength
measurements to further test the nature of source C1.

{\it Conclusion.} ---
The SNR-MC interacting systems are believed to be ideal targets to probe acceleration of 
hadronic CRs by SNR shocks. Some of these systems exhibit characteristic $\pi^0$-decay
bumps in their $\gamma$-ray spectra, strengthening the evidence that SNRs are one class
of sources of Galactic CRs. In this work, we carry out detailed study of the morphology 
and spectrum of very high energy $\gamma$-ray emission from such an example, the region of 
SNR IC 443, with the LHAASO data. Two sources have been resolved in the data, one is a point 
source (C0) coincides with the compact source detected by Fermi-LAT, MAGIC, VERITAS, and 
HAWC which has both the interaction with MCs and the $\pi^0$-decay bump, and the other is 
an extended source (C1) coincides with the newly reported Fermi-LAT source 2FGES J0618.3+2227. 
The spectrum of C0 is well described by a PL shape without significant cutoff, and the spectrum 
of C1 can be described by an ECPL shape. The LHAASO SED of source C0 connects smoothly with 
the Fermi-LAT compact source, and the wide-band $\gamma$-ray SED can be well modelled with 
a hadronic scenario. We derive the $95\%$ lower limit of the cutoff momentum of protons for 
source C0 to be $\sim300$ TeV, providing compelling evidence that the SNR shock can accelerate 
protons to sub-PeV energies. The location and extension of source C1 are consistent with the 
Fermi-LAT extended one, and the SEDs are also consistent with each other at overlapping energies.
Distributed molecular gas exist in the sky region of the source, indicating that the $\gamma$-ray 
emission may have a hadronic origin. The proton spectrum to account for the wide-band SED of C1 is
different from that of C0, which may be interpreted as a propagation effect of escaping protons.
Alternatively, a leptonic scenario can also explain the $\gamma$-ray emission of C1. 

The data that support the findings of this article are openly
available\footnote{https://www.nhepsdc.cn/resource/astro/lhaaso/20260221095403}.

\acknowledgments
We would like to thank all staff members who work at the LHAASO site 4400 m above
the sea level year-round to maintain the detector and keep the water recycling system, 
electricity power supply, and other components of the experiment operating smoothly.
We are grateful to Chengdu Management Committee of Tianfu New Area for the constant 
financial support for research with LHAASO data. We appreciate the computing and data 
service support provided by the National High Energy Physics Data Center. 
This work is supported by the following grants: the National Natural Science Foundation 
of China (Nos. 12220101003,  12322302, 12573053), the CAS Project for Young Scientists 
in Basic Research (No. YSBR-061), the Natural Science Foundation for General Program of 
Jiangsu Province of China (No. BK20242114), and the Jiangsu Provincial Excellent 
Postdoctoral Program (No. 2022ZB472), the National Science and Technology Development 
Agency (NSTDA) of Thailand, and the National Research Council of Thailand (NRCT) under 
the High-Potential Research Team Grant Program (N42A650868).
This study made use of the data from the Milky Way Imaging Scroll Painting (MWISP) 
project, which is a multi- line survey in 12CO/13CO/C18O along the northern galactic 
plane with the PMO-13.7m telescope. We are grateful to all the members of the MWISP 
working group, particularly the staff members at the PMO-13.7m telescope, for their 
long-term support. MWISP was sponsored by National Key Research and Development Program 
of China with grants 2023YFA1608000, 2017YFA0402701, and by CAS Key Research Program 
of Frontier Sciences with grant QYZDJ-SSW-SLH047.

\bibliographystyle{apsrev}
\bibliography{main}

\clearpage


\onecolumngrid

\setcounter{figure}{0}
\renewcommand\thefigure{S\arabic{figure}}
\setcounter{table}{0}
\renewcommand\thetable{S\arabic{table}}



\begin{center}
    {\Large Supplemental Material of ``Evidence of cosmic-ray acceleration up to sub-PeV energies in the supernova remnant IC 443''}\\
    (The LHAASO collaboration)\\
\end{center}

\section{ROI of the analysis}
\begin{figure}[!htb]
\centering
\includegraphics[width=0.49\textwidth]{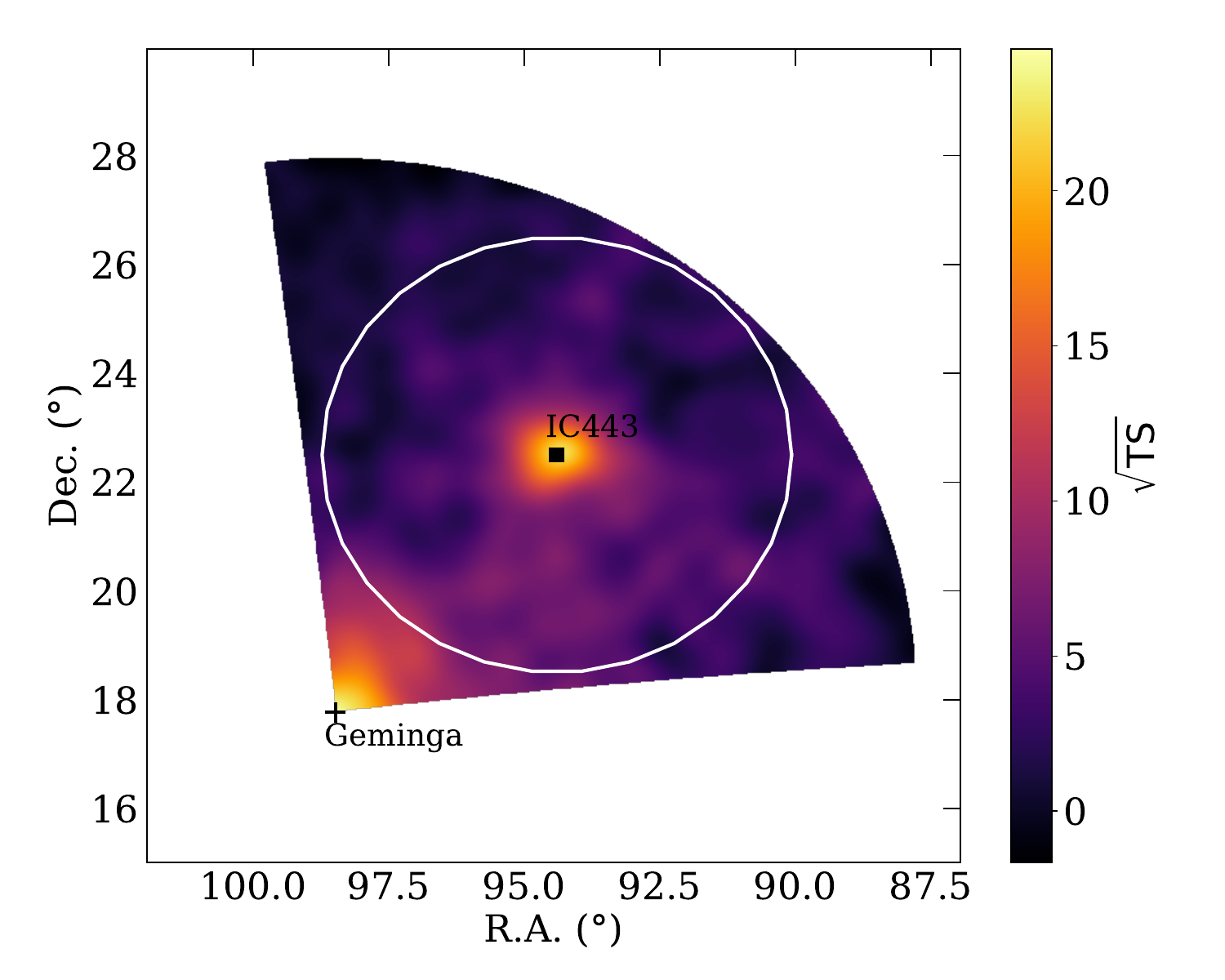}
\includegraphics[width=0.42\textwidth]{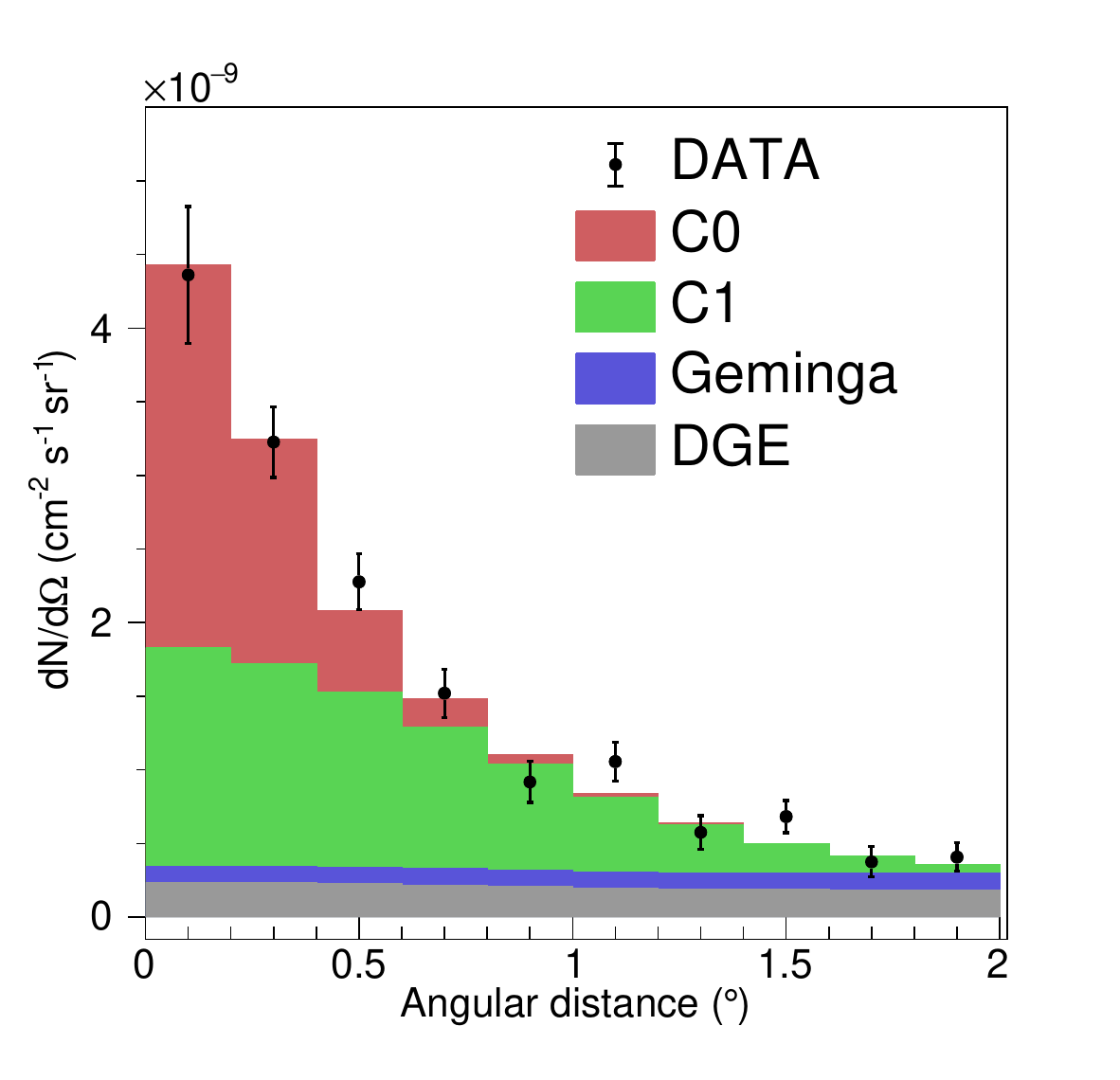}
\caption{Left: the fan-shaped region of interest (ROI) of this analysis, with the white circle denoting a $4^{\circ}$ radius centered on IC 443.
Right: the one-dimensional radial distribution of the integrated $\gamma$-ray fluxes above 1 TeV for different components, centered at the midpoint between C0 and C1.}
\label{fig:ROI}
\end{figure}

The ROI of the analysis is a fan-shaped region centered at Geminga pulsar with an opening angle of $90^{\circ}$, as shown in the left panel of Fig.~\ref{fig:ROI}. The radial distributions of C0 and C1 differ significantly from those of the diffuse emission and the Geminga halo, as illustrated in the right panel of Fig.~\ref{fig:ROI}. Since C0 and C1 are about 6 degrees away from Geminga, the radial profile of Geminga around C0 and C1 is relatively flat. Besides, enlarging or reducing the opening angle of fan-shaped region by $20^{\circ}$ have been tested, and the results are almost unchanged.

\section{Systematic uncertainties}
\begin{figure*}[!htb]
\centering
\includegraphics[width=0.48\textwidth]{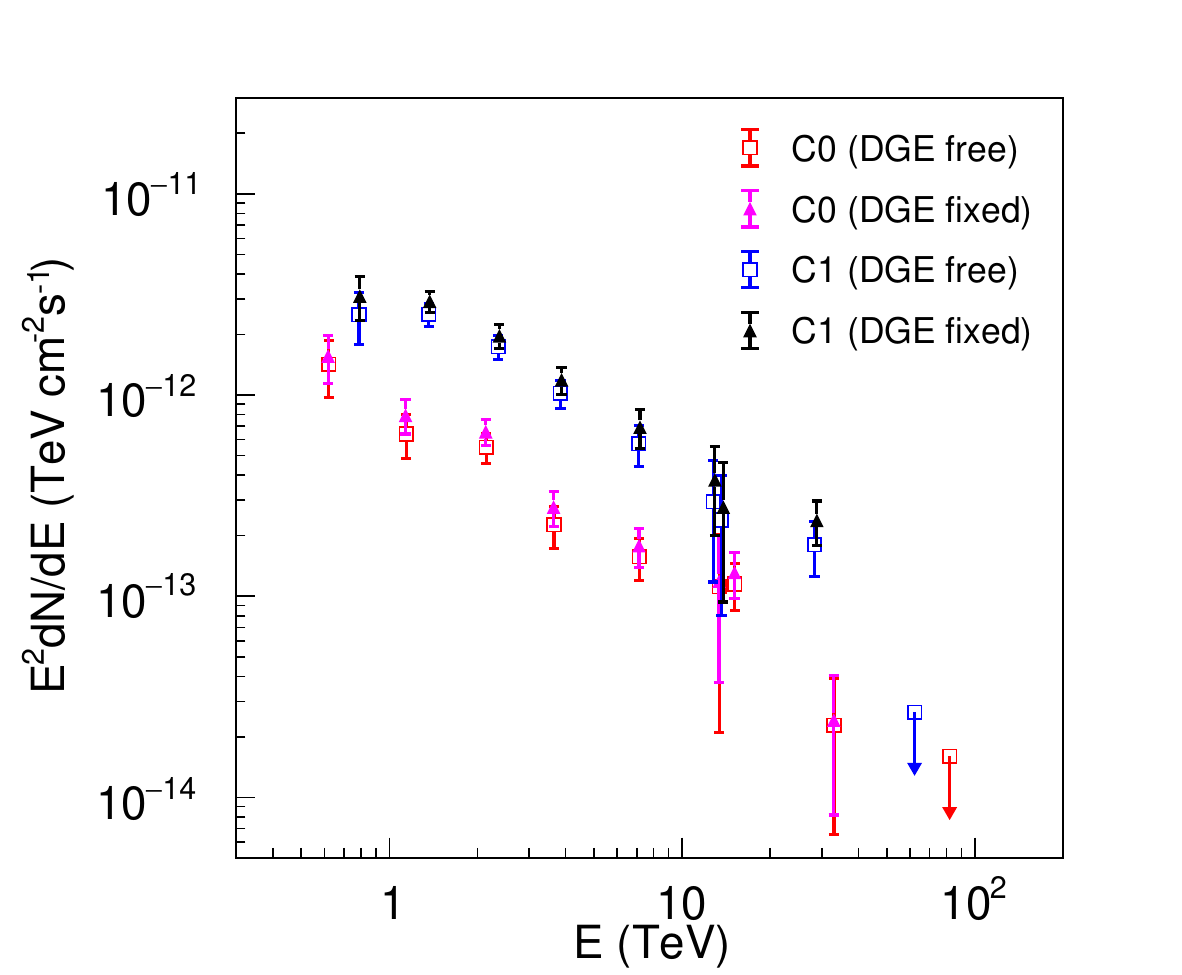}
\includegraphics[width=0.48\textwidth]{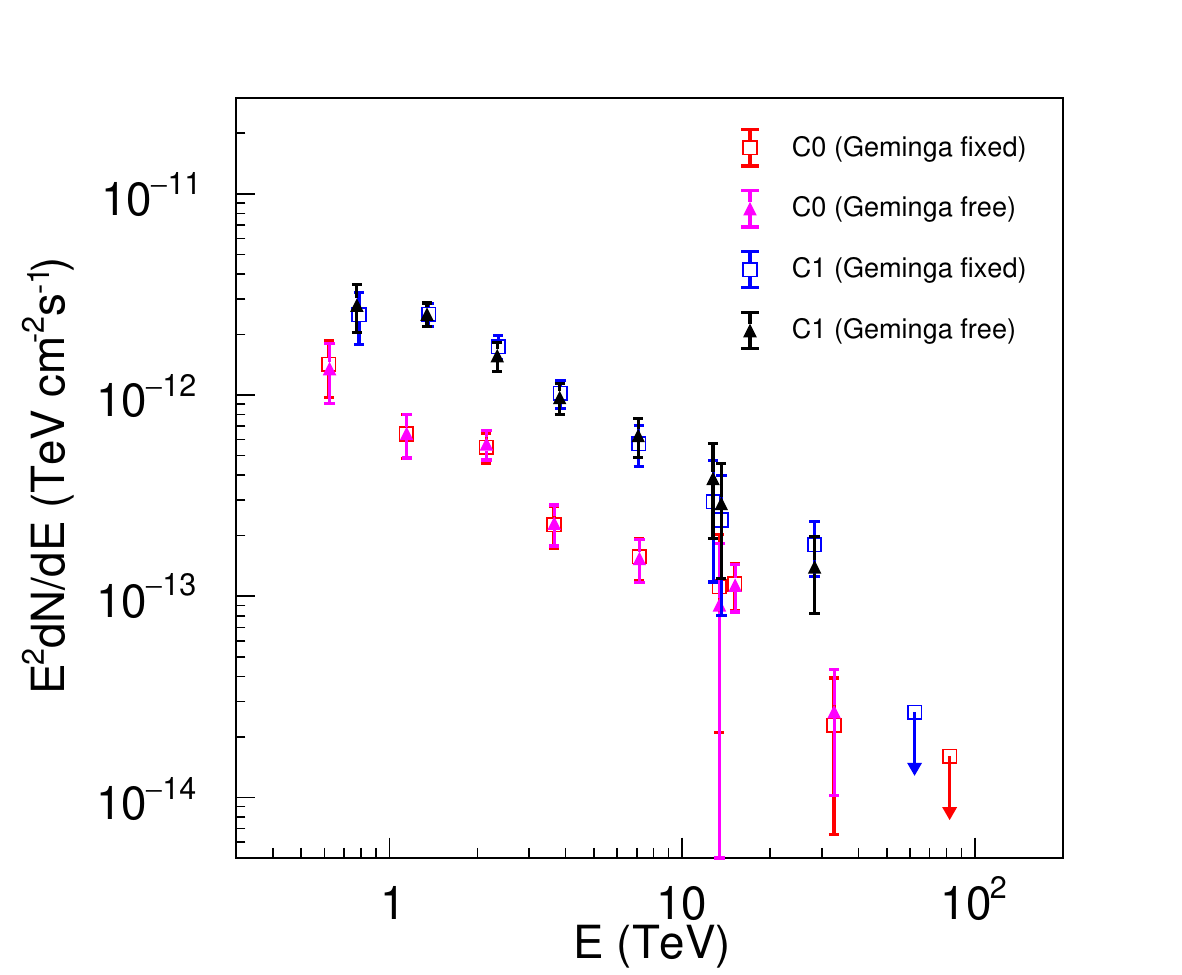}
\caption{Impact of the diffuse emission (left panel) and Geminga halo (right panel) 
on the SEDs of sources C0 and C1.}
\label{fig:SED_sys}
\end{figure*}

The diffuse emission may have some impacts on the measurements of IC 443. We compare the results 
for two assumptions of the diffuse emission, with fixed flux from the outer Galactic plane
\cite{2025PhRvL.134h1002C} and a free flux normalization. The results are shown in the left panel of
Fig.~\ref{fig:SED_sys}. The TS value decreases by about 15 when we fix the diffuse flux normalization, 
and the flux from the diffuse component is higher than the average flux from the outer Galactic plane
\cite{2025PhRvL.134h1002C}. However, the results of IC 443 are affected very slightly by different 
diffuse emission assumptions. Results of all parameters of sources C0 and C1 are consistent within
statistical errors. A different diffuse emission template from the gas surveys (HI, H$_2$, and HII) 
has been tested, and very minor differences on the results of C0 and C1 have been found (see Table
\ref{tab:sys}).

The Geminga halo extends to the region of IC 443 and may affect the analysis of IC 443. 
The Geminga halo shows asymmetric morphology which has been used as the benchmark of this
analysis. As a test, using the symmetric morphology for Geminga in this analysis decreases the 
total TS value of the two sources C0 and C1 by only 1.2, suggesting that no strong asymmetry 
exists in the ROI. The fitting results of C0 and C1 are consistent with the benchmark setting 
(Table \ref{tab:sys}).
In addition, for the benchmark setting of this work, we assume that the energy-dependence of the 
extensions of Geminga halo follows a power-law form. To address this impact, we leave the extension
parameter ($\theta_d$) of Geminga free in each energy bin, and re-derive the fluxes of IC 443. 
The differences in the resulting SEDs are shown in the right panel of Fig.~\ref{fig:SED_sys}. 
It is shown that for $E<20$ TeV the results are in good agreement with each other, and slight 
differences exist for higher energies. 

Other systematic uncertainties on the absolute flux measurements are estimated to be
about $9.6\%$ for WCDA \cite{2023Sci...380.1390L} and $7\%$ for KM2A \cite{2021ChPhC..45b5002A}, 
due mainly to various kinds of model assumptions of the Monte Carlo simulation. All the systematic
uncertainties are added in quadrature to get the total systematic uncertainties of the flux measurements.

\begin{table*}[!htb]
\centering
\footnotesize
\begin{tabular}{l|c|c|c|c|c|c|c|c|c}
\hline
Model & $\Delta$TS & R.A. & Dec. & $R_{39}$ & Flux @ 3 TeV & $\alpha$ & $E_{\rm{cut}}$ & $\Phi_0$ (DGE) & Comment\\ 
 & & ($^\circ$) & ($^\circ$) & ($^\circ$) & ($10^{-14}$ TeV$^{-1}$cm$^{-2}$s$^{-1}$) &  & (TeV) & (TeV$^{-1}$cm$^{-2}$s$^{-1}$) & \\ \hline 
C0 &  & 94.27$\pm$0.03 & 22.44$\pm$0.02 & 0.01$\pm$0.20 & 3.51$\pm$0.43 & 2.95$\pm$0.07 & -- & 5.35$\pm$0.37 & \multirow{2}{*}{Benchmark}  \\ 
C1 & & 94.45$\pm$0.07 & 22.61$\pm$0.06 & 0.67$\pm$0.07 & 17.20$\pm$2.74 & 2.53$\pm$0.14 & 19.65$\pm$8.67 &   (Planck) & \\
\hline
C0 & \multirow{2}{*}{$-15.3$} & 94.28$\pm$0.03 & 22.45$\pm$0.02 & 0.07$\pm$0.21 & 4.09$\pm$0.39 & 2.97$\pm$0.06 & -- &  3.84 & \multirow{2}{*}{DGE fixed}\\
C1 & & 94.43$\pm$0.06 & 22.59$\pm$0.07 & 0.82$\pm$0.08 & 19.98 $\pm$2.15 & 2.53$\pm$0.10  & 22.01 $\pm$ 8.11  & (Planck) & \\
\hline
C0 & \multirow{2}{*}{$+0.6$} & 94.27$\pm$0.03 & 22.44$\pm$0.02 & 0.01$\pm$0.20 & 3.45$\pm$0.40 & 2.95$\pm$0.07 & -- & 6.32$\pm$ 0.44 & DGE template \\
C1 & & 94.42$\pm$0.06 & 22.60$\pm$0.06 & 0.66$\pm$0.06 & 16.98 $\pm$2.18 & 2.52$\pm$0.12  & 22.01 $\pm$ 6.86  & (gas survey) & from gas survey\\
\hline
C0 & \multirow{2}{*}{$-1.2$} & 94.27$\pm$0.03 & 22.44$\pm$0.02 & 0.01$\pm$0.20 & 3.48$\pm$0.34 & 2.96$\pm$0.07 & -- & 6.50$\pm$0.45 & Symmetric Geminga\\
C1 & & 94.42$\pm$0.07 & 22.59$\pm$0.06 & 0.67$\pm$0.05 & 16.61 $\pm$2.11 & 2.49$\pm$0.12  & 21.07 $\pm$ 6.80  &  (Planck) & in the ROI \\
\hline
\end{tabular}
\begin{tablenotes}
\footnotesize
\centering
\item[1]* DGE spectrum: $\Phi=\Phi_0\cdot10^{-14}(E/10 ~{\rm TeV})^{-2.72}\left[1+\left(E/27.86~{\rm TeV}\right)^{5}\right]^{(2.72-2.92)/5}$ \cite{2025PhRvL.134h1002C}.
\end{tablenotes}
\caption{Impacts on the results of IC 443 for different settings.}
\label{tab:sys}
\end{table*}

\section{LHAASO fluxes of sources C0 and C1}
Table \ref{tab:flux} give the measured fluxes of source C0 and C1 by LHAASO, with statistical and systematic uncertainties.

\begin{table*}[!htb]
\centering
\begin{tabular}{cc|cc} \hline\hline
\multicolumn{2}{c|}{C0} & \multicolumn{2}{c}{C1} \\ \hline
$E$ & $E^2dN/dE \pm \sigma_{\rm stat} \pm \sigma_{\rm sys} $ & $E$ & $E^2dN/dE \pm \sigma_{\rm stat} \pm \sigma_{\rm sys} $ \\
(TeV) & (TeV~cm$^{-2}$~s$^{-1}$) & (TeV) & (TeV~cm$^{-2}$~s$^{-1}$) \\ \hline
0.62 & $(1.42 \pm 0.45 \pm 0.21) \times 10^{-12}$ & 0.79 & $(2.51 \pm 0.72 \pm 0.70) \times 10^{-12}$ \\
1.14 & $(6.40 \pm 1.57 \pm 1.65) \times 10^{-13}$ & 1.36 & $(2.52 \pm 0.32 \pm 0.47) \times 10^{-12}$ \\
2.14 & $(5.51 \pm 0.96 \pm 1.20) \times 10^{-13}$ & 2.36 & $(1.74 \pm 0.24 \pm 0.32) \times 10^{-12}$ \\
3.65 & $(2.26 \pm 0.53 \pm 0.56) \times 10^{-13}$ & 3.84 & $(1.02 \pm 0.16 \pm 0.20) \times 10^{-12}$ \\
7.14 & $(1.57 \pm 0.37 \pm 0.26) \times 10^{-13}$ & 7.11 & $(5.74 \pm 1.31 \pm 1.43) \times 10^{-13}$ \\
15.15 & $(1.15 \pm 0.30 \pm 0.20) \times 10^{-13}$ & 13.63 & $(2.39 \pm 1.59 \pm 0.67) \times 10^{-13}$ \\
13.40 & $(1.12 \pm 0.91 \pm 0.24) \times 10^{-13}$ & 12.78 & $(2.95 \pm 1.77 \pm 1.25) \times 10^{-13}$ \\
33.04 & $(2.29 \pm 1.64 \pm 0.43) \times 10^{-14}$ & 28.33 & $(1.80 \pm 0.55 \pm 0.72) \times 10^{-13}$ \\
82.20 & $<1.60\times 10^{-14}$  & 62.25 & $<2.66\times 10^{-14}$ \\
\hline \hline
\end{tabular}
\caption{Fluxes of LHAASO sources C0 and C1, with $1\sigma$ statistical and systematic uncertainties.}
\label{tab:flux}
\end{table*}

\section{Fermi-LAT analysis}

\begin{table}[!htb]
  \centering
  \begin{tabular}{ccc}
    \hline\hline 
    $\log(E/\mathrm{MeV})$ & $\phi_{\rm 4FGL}$ ($\mathrm{MeV^{-1}cm^{-2}s^{-1}}$)   & $\phi_{\rm 2FGES}$ ($\mathrm{MeV^{-1}cm^{-2}s^{-1}}$) \\
    \hline 
    1.78--1.95 & $< 2.88 \times 10^{-9}$ & $< 2.55 \times 10^{-9}$ \\
    1.95--2.12 & $(1.42 \pm 0.44) \times 10^{-9}$ & $< 1.40 \times 10^{-9}$ \\
    2.12--2.28 & $(1.16 \pm 0.14) \times 10^{-9}$ & $< 3.49 \times 10^{-10}$ \\
    2.28--2.45 & $(7.28 \pm 0.30) \times 10^{-10}$ & $< 5.10 \times 10^{-11}$ \\
    2.45--2.62 & $(4.03 \pm 0.15) \times 10^{-10}$ & $< 2.86 \times 10^{-11}$ \\
    2.62--2.79 & $(2.15 \pm 0.05) \times 10^{-10}$ & $< 1.06 \times 10^{-11}$ \\
    2.79--2.96 & $(1.04 \pm 0.02) \times 10^{-10}$ & $(4.83 \pm 2.32) \times 10^{-12}$ \\
    2.96--3.13 & $(5.17 \pm 0.09) \times 10^{-11}$ & $< 2.94 \times 10^{-12}$ \\
    3.13--3.29 & $(2.28 \pm 0.04) \times 10^{-11}$ & $(8.06 \pm 4.13) \times 10^{-13}$ \\
    3.29--3.47 & $(9.92 \pm 0.19) \times 10^{-12}$ & $< 5.60 \times 10^{-13}$ \\
    3.47--3.64 & $(4.33 \pm 0.10) \times 10^{-12}$ & $(2.00 \pm 0.84) \times 10^{-13}$ \\
    3.64--3.80 & $(1.75 \pm 0.05) \times 10^{-12}$ & $(1.72 \pm 0.41) \times 10^{-13}$ \\
    3.80--3.97 & $(6.84 \pm 0.24) \times 10^{-13}$ & $(6.51 \pm 1.91) \times 10^{-14}$ \\
    3.97--4.14 & $(2.63 \pm 0.12) \times 10^{-13}$ & $(2.03 \pm 0.89) \times 10^{-14}$ \\
    4.14--4.31 & $(1.10 \pm 0.06) \times 10^{-13}$ & $(1.15 \pm 0.44) \times 10^{-14}$ \\
    4.31--4.48 & $(3.65 \pm 0.29) \times 10^{-14}$ & $(6.11 \pm 2.27) \times 10^{-15}$ \\
    4.48--4.65 & $(1.38 \pm 0.15) \times 10^{-14}$ & $(3.24 \pm 1.17) \times 10^{-15}$ \\
    4.65--4.82 & $(4.38 \pm 0.71) \times 10^{-15}$ & $(1.63 \pm 0.61) \times 10^{-15}$ \\
    4.82--4.99 & $(1.11 \pm 0.31) \times 10^{-15}$ & $(7.08 \pm 3.17) \times 10^{-16}$ \\
    4.99--5.16 & $(2.78 \pm 1.48) \times 10^{-16}$ & $(3.59 \pm 1.76) \times 10^{-16}$ \\
    5.16--5.32 & $(1.29 \pm 0.69) \times 10^{-16}$ & $< 2.68 \times 10^{-16}$ \\
    5.32--5.49 & $(1.02 \pm 0.50) \times 10^{-16}$ & $(9.70 \pm 5.44) \times 10^{-17}$ \\
    5.49--5.66 & $(3.20 \pm 2.19) \times 10^{-17}$ & $< 3.86 \times 10^{-17}$ \\
    5.66--5.83 & $< 1.26 \times 10^{-17}$ & $< 1.64 \times 10^{-17}$ \\
    5.83--6.00 & $< 1.85 \times 10^{-17}$ & $< 8.59 \times 10^{-18}$ \\
    \hline\hline 
  \end{tabular}
  \caption{Fluxes with 1$\sigma$ uncertainties for sources 4FGL J0617.2+2234e and 2FGES J0618.3+2227 measured by Fermi-LAT.}
  \label{tab:flux_fermi}
\end{table}

In this work, we re-analyze the Fermi-LAT data of the IC 433 region with larger data set. 
The newest reconstructed P8R3 SOURCE Fermi-LAT
data\footnote{https://fermi.gsfc.nasa.gov/ssc/data/access/}
are used in this analysis \cite{2013arXiv1303.3514A}. We select the data recorded from August 4, 
2008 to February 5, 2025, 870 weeks in total. To verify the $\pi^0$-decay bump observed from IC 
443, photons with energies down to 60 MeV are selected. To suppress the contamination from 
$\gamma$-rays generated by cosmic ray interactions in the upper layers of the atmosphere, 
photons collected at zenith angles larger than 90$^\circ$ are removed. Moreover, we filter the 
data using the specification {\tt{(DATA$\_$QUAL$>$0)~$\&\&$~(LAT$\_$CONFIG==1)}} to select 
good time intervals in which the satellite was working in the standard data taking mode 
and the data quality is good. We bin the data, from 60 MeV to 1 TeV, into 50 logarithmically 
distributed energy bins and $200\times200$ spatial bins with size 0.1$^\circ$ centered at 
IC 443. We employ the binned likelihood analysis method to analyze the data with Fermitools 
{\tt version 2.2.0}\footnote{https://fermi.gsfc.nasa.gov/ssc/data/analysis/documentation/}. 
The instrument response function (IRF) adopted is {\tt P8R3\_SOURCE\_V3}. 
The energy dispersion may be important for the analysis with low energy data, and is taken 
into account in the likelihood fitting. For the diffuse background 
emissions, we take the Galactic diffuse model {\tt gll\_iem\_v07.fits} and the isotropic 
background spectrum {\tt iso\_P8R3\_SOURCE\_V3\_v1.txt} as recommended by the Fermi-LAT
collaboration\footnote{http://fermi.gsfc.nasa.gov/ssc/data/access/lat/BackgroundModels.html}. 
The source model XML file is generated using the user contributed tool 
{\tt {make4FGLxml.py}}\footnote{http://fermi.gsfc.nasa.gov/ssc/data/analysis/user/} based on the 
4FGL source catalog \cite{2020ApJS..247...33A,2023arXiv230712546B}, including the new extended 
source with the same coordinate and extension as reported in Ref.~\citep{2024arXiv241107162A} 
and a power-law spectrum. We first make a broadband fitting to get the best fitted parameters 
for sources in the region of interest. Due to the large PSF of Fermi-LAT at low energies, 
we re-select data from 60 MeV to about 200 MeV with an extra cut of {\tt PSF3} to reduce the 
degeneracy between IC 443 and the new extended source. Further, for this data set, IRF 
{\tt P8R3\_SOURCE\_V3::PSF3} is adopted and the isotropic background spectrum 
{\tt iso\_P8R3\_SOURCE\_PSF3\_V3\_v1.txt} is used to match the data cut. Then we extract 
the SEDs, 60 MeV to about 200 MeV from the {\tt PSF3} data set and above about 200 MeV from 
the original data set, for IC 443 and the new extended source with other point sources parameters 
fixed to the best-fitting values obtained above. The obtained fluxes are reported in 
Table~\ref{tab:flux_fermi}. Compared with previous Fermi-LAT analyses, our results for the 
large extended source are well consistent with those given in Ref.~\citep{2024arXiv241107162A}.
For the compact source, our derived fluxes agree well with the spectrum in the 4FGL catalog
\cite{2020ApJS..247...33A}, but are slightly lower than those reported in Ref.~\cite{2013Sci...339..807A}.
Such differences might be attributed to the data processed with state-of-the-art event reconstruction 
in this work.


\end{document}